# A Systematic Search for Supersoft X-Ray Sources in the ROSAT All-Sky Survey

J. Greiner

Max-Planck-Institut für Extraterrestrische Physik, 85740 Garching, Germany

**Abstract.** We have conducted a systematic search for supersoft X-ray sources using the ROSAT all-sky survey data. With the optical identification of the selected sources being almost complete, we discuss the statistics of the various source classes and their observability. Besides supersoft close binary sources this search also can be used to estimate the number of isolated neutron stars in the Galaxy, such as those described by Stocke et al. (1995) and Walter et al. (1996).

## 1 Introduction

Supersoft X-ray sources (SSS) are characterized by very soft X-ray radiation (most of the X-ray emission below 0.5 keV) of high luminosity. It is generally accepted that SSS involve steady nuclear burning on the surface of an accreting WD (van den Heuvel et al. 1992). ROSAT observations established SSS as a distinct class with somewhat more than 30 members known at present. While 16 sources belong to M 31, and a dozen to the Magellanic Clouds, only very few have been found in the Galaxy.

A simple scaling from the brightest supersoft X-ray sources in the Magellanic Clouds to galactic distances already shows that there are no such sources in our immediate neighbourhood. If CAL 83, the SSS prototype (Long et al. 1981), were at 1 kpc distance, we would expect a ROSAT PSPC countrate of 3000 cts/sec, much larger than the PSPC threshold intensity. Any such unabsorbed source in the solar neighbourhood would have forced the PSPC to switch off and it is safe to say that no such source has been detected in the ROSAT all-sky survey.

However, the intrinsic source luminosities are observed to vary from source to source by a factor of about 50, the mean temperatures of the WDs could be systematically lower in the Galaxy as compared to the Magellanic Clouds, and a moderate absorbing column could further dim the sources. We therefore have undertaken a systematic search of the ROSAT PSPC all-sky survey data for supersoft X-ray sources with emphasis on the galactic population.

## 2 Selection Criteria

We started with the list containing all detected X-ray sources of the all-sky survey in its version of March 1991 which contained about 100.000 sources (including detections of identical sources in overlapping strips). We applied a hardness ratio criterion $HR1 + \sigma_{HR1} \leq -0.80$ which is fulfilled by 304 sources. The hardness ratio HR1 is defined as the normalized count difference ($N_{52-201} - N_{11-41})/(N_{11-41} + N_{52-201}$), where $N_{a-b}$ denotes the number of counts in the PSPC between channel a and channel b. After merging double and multiple detections from different strips at high ecliptic latitudes we are left with 165 sources. We have taken the results of each source from that strip in which the source has the largest distance to either edge of the strip. Finally, images in different energy bands were visually inspected to check for possible false detections, and to reduce spurious sources in bright supernova remnants (like Vela/Puppis). The final number of selected sources is 143.

This kind of hardness ratio selection has the implication of an implicit intensity selection. First, for decreasing brightness the error in the hardness ratio increases, and all sources with $\sigma_{HR1} \geq 0.2$ are exluded. Second, it also cuts away specific sources (like G or K stars) which are detected at high signal to noise ratio. This is due to the fact that these objects exhibit a faint hard component, and this is detected only for bright sources.

Finally we should note that the input list has been produced from source detections on the individual strips. At high ecliptic latitudes these strips overlap, and with the "survey II" processing which is done on adjacent sky fields instead of strips (and which will becaome available soon) a considerable improvement of the source detection and parameter estimation is possible.

## 3    Optical Identification and Source Statistics

A correlation with the Simbad database revealed probable source identifications for a total of 48 sources. Most of these identifications were single white dwarfs. Since we were not interested in discovering further WDs, and many of these WDs have been detected also in the extreme-ultraviolet sensitive Wide Field Camera (WFC), we have used the WFC source list to remove all sources from our master list which have been seen in the WFC (total number of 53). These sources were not considered for our own follow-up identification, but were left by purpose for the WFC consortium (and most of these were later identified as WDs).

The remaining sources have been observed optically by the author (mainly northern hemisphere) and the group of K. Beuermann (southern hemisphere). The majority of X-ray sources turned out to be magnetic cataclysmic variables, predominantly polars. The only galactic supersoft source (close binary) in this sample is RX J0019.8+2156 (Beuermann et al. 1995, Greiner & Wenzel 1995) while the other two are LMC/SMC sources (1E 0035.4–7230 and RX J0439.8–6809). Tab. 1 shows a breakdown of the sources on the individual source classes.

A few notes should be added to Tab. 1. The G and K star identifications (IDs) are probably systems with a WD in a close orbit, so after spectroscopic follow-up observations these two IDs will certainly be replaced. There is one additional object of presumably this type among the unidentified sources which also has a G star at the X-ray position. The only active galactic nuclei in our sample is the narrow-line Seyfert 1 (NLSy1) galaxy WPVS007, and many more are expected at somewhat harder hardness ratios (Greiner et al. 1996).

**Table 1.** The distribution among different classes of objects of supersoft X-ray sources which satisfy HR1 + $\sigma_{\mathrm{HR1}} \leq -0.80$. PN means planetary nebula, the other abbreviations are explained in the text.

| Class type | Simbad correlation | newly identified | total |
|---|---|---|---|
| SSS | 0 | 3 | 3 |
| PN | 2 | 0 | 2 |
| PG1159 stars | 0 | 1 | 1 |
| symbiotics | 2 | 1 | 3 |
| single WDs | 32 | 66 | 98 |
| magnetic CVs | 5 | 12 | 17 |
| NLSy1 | 1 | 0 | 1 |
| B stars | 4 | 0 | 4 |
| G stars | 1 | 0 | 1 |
| K stars | 1 | 0 | 1 |
| unidentified | | | 12 |

## 4  Discussion

In total, only seven out of the more than 30 known supersoft sources are members of our sample, namely the above listed RX J0019.8+2108, 1E 0035.4–7230 and RX J0439.8–6809 plus the planetary nebula RX J0058.6–7146 (N 67) plus the three symbiotics RX J0048.4–7332 (SMC3), RR Tel and AG Dra. This demonstrates the "conservative" approach of our selection criterium. In fact, it is more effective for low-temperature sources like single white dwarfs or PG1159 stars. The above listed sources are the softest among the known population of SSS. As soon as a source has a temperature higher than about 40 eV and the PSPC detects a significant number of photons above channel 41 (0.4 keV) than the HR1 increases and shifts the source out of our hardness ratio window. That is, our sampling of supersoft sources is complete only for temperatures below about 30 eV.

The brightest of the non-identified sources has a PSPC countrate of 0.37 cts/s. Above this intensity the identification is complete. This implies that any galactic source radiating at the Eddington rate with a temperature in the 20–40 eV range would have been detected above a galactic latitude bII>12°. Relaxing the luminosity constraint by a factor of 10 increases the latitude limit only slightly to bII>15°.

A different approach in searching for supersoft sources is to look specifically for strongly absorbed sources in the galactic plane. Their energy distribution then is expected to be sharply peaked around 0.5–0.7 keV with higher energy photons missing due to the spectrum of the source, and lower energy photons absorbed. Such a source would have a relatively hard HR1 but a very soft HR2. Indeed, these criteria have led to the discovery of RX J0925.7–4758 (Motch et al. 1994). From the results of the galactic plane survey identification these authors concluded that though sources radiating at the Eddington limit with a temperature below ∼20 eV may be hidden in the galactic plane, sources at bII > 5–10° with temperatures above 40 eV and $4\times10^{37}$ erg/s bolometric luminosity can be excluded. This supplements our finding for lower temperature sources.

Another possible class of soft X-ray sources are old, isolated neutron stars (IONS) accreting from the interstellar matter. Two sources have been proposed to be IONS, one of those (RX J1856.5–3754) is a bright source seen in the ROSAT all-sky survey (Walter et al. 1996). This source has been reported to have a best-fit blackbody temperature of below 60 eV. However, due to the distinct emission above 0.4 keV the hardness ratio is HR1=–0.29, far outside our selected range.

Such IONS are not expected to be easily visible in the optical range. All the unidentified sources of our sample have optical counterpart candidates brighter than 20th magnitude inside their corresponding error box. We therefore think that none of these sources might be another IONS candidate. This makes our search complete down to a limiting countrate of 0.1 cts/sec.

Since the temperature of the emitted radiation (assumed to be blackbody) is primarily determined by the ratio of accretion rate $\dot{M}$ to the fractional accreting area f (kT = 20 $(\dot{M}/f)^{1/4}$ eV with $\dot{M}$ in units of $10^{10}$ g/s, Blaes & Madau 1993), we can determine the range of accretion rates for which our search for supersoft sources is sensitive. Assuming isotropic accretion for simplicity, sources with accretion rates above $50\times10^{10}$ g/s are too hot for our search (have a hardness ratio greater than –0.8), while below $0.05\times10^{10}$ g/s the temperature and luminosity get too low to be detectable beyond 20 pc.

*Acknowledgement:* JG is extremely grateful to T. Fleming for the frequent information on the status of single WD identifications. JG is supported by the Deutsche Agentur für Raumfahrtangelegenheiten (DARA) GmbH under contract FKZ 50 OR 9201. The *ROSAT* project is supported by the German Bundesministerium für Bildung, Forschung, Wissenschaft und Technologie (BMBW/DARA) and the Max-Planck-Society. This research has made use of the Simbad database, operated at CDS, Strasbourg, France.